\documentclass[letterpaper]{article} 
\usepackage{aaai2026}  
\usepackage{times}  
\usepackage{helvet}  
\usepackage{courier}  
\usepackage[hyphens]{url}  
\usepackage{graphicx} 
\usepackage{natbib}  
\usepackage{caption} 
\usepackage{subcaption} 
\usepackage{tabularx}
\usepackage{booktabs}
\usepackage{array,makecell}
\newcolumntype{C}[1]{>{\centering\arraybackslash}m{#1}}
\newcolumntype{L}[1]{>{\raggedright\arraybackslash}m{#1}} 

\usepackage{algorithm}
\usepackage{algorithmic}

\usepackage{newfloat}
\usepackage{listings}
\DeclareCaptionStyle{ruled}{labelfont=normalfont,labelsep=colon,strut=off} 
\floatstyle{ruled}
\newfloat{listing}{tb}{lst}{}
\floatname{listing}{Listing}
\title{Towards Human-AI Accessibility Mapping in India: VLM-Guided Annotations and POI-Centric Analysis in Chandigarh}

\author{
Varchita Lalwani\textsuperscript{\rm 1},
Utkarsh Agarwal\textsuperscript{\rm 1},
Michael Saugstad\textsuperscript{\rm 2},
Manish Kumar\textsuperscript{\rm 3},\\
Jon E. Froehlich\textsuperscript{\rm 2},
Anupam Sobti\textsuperscript{\rm 1}
}

\affiliations{
\textsuperscript{\rm 1}Plaksha University, Mohali, India\\
\textsuperscript{\rm 2}University of Washington, Seattle, WA, USA\\
\textsuperscript{\rm 3}Indian Institute of Technology Delhi, New Delhi, India\\

}

\usepackage{bibentry}
\begin{document}
\maketitle

\begin{figure*}[!t]
\centering
\begin{subfigure}[t]{0.2\textwidth}
    \centering
    \includegraphics[width=1\linewidth]{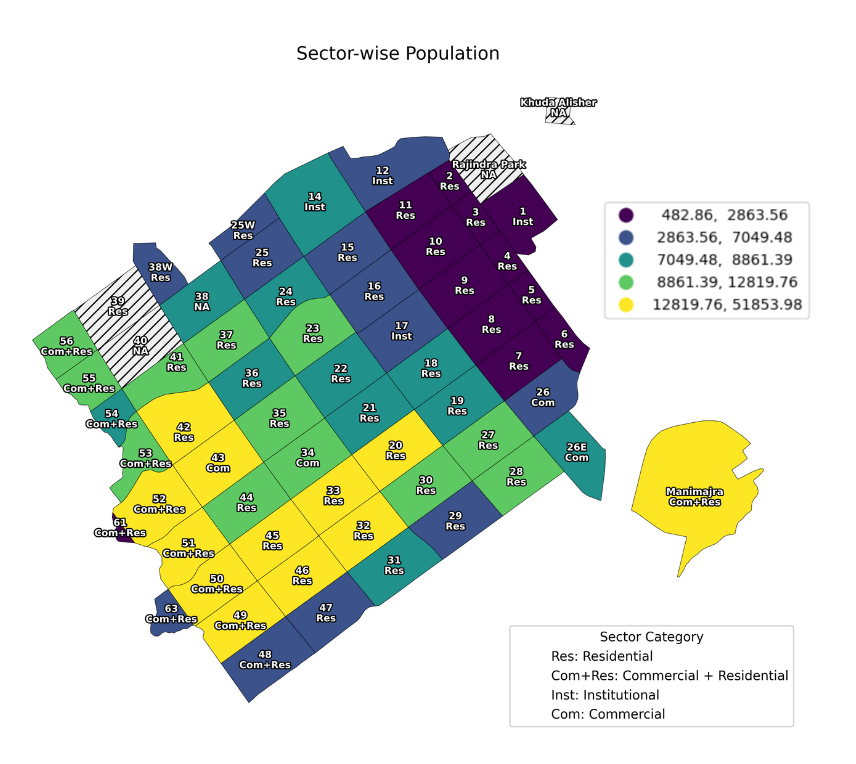}
    \caption{Sector-wise population heatmap of Chandigarh showing relative population density across sectors.}
    \label{fig:population_heatmap}
\end{subfigure}
%
\begin{subfigure}[t]{0.2\textwidth}
  \centering
  \includegraphics[width=\linewidth]{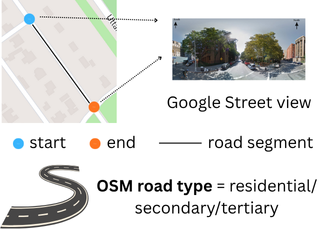}
  \caption{Input: OSM road type + start/end Street View panoramas}
  \label{fig:a}
\end{subfigure}
\begin{subfigure}[t]{0.2\textwidth}
  \centering
  \includegraphics[width=1\linewidth]{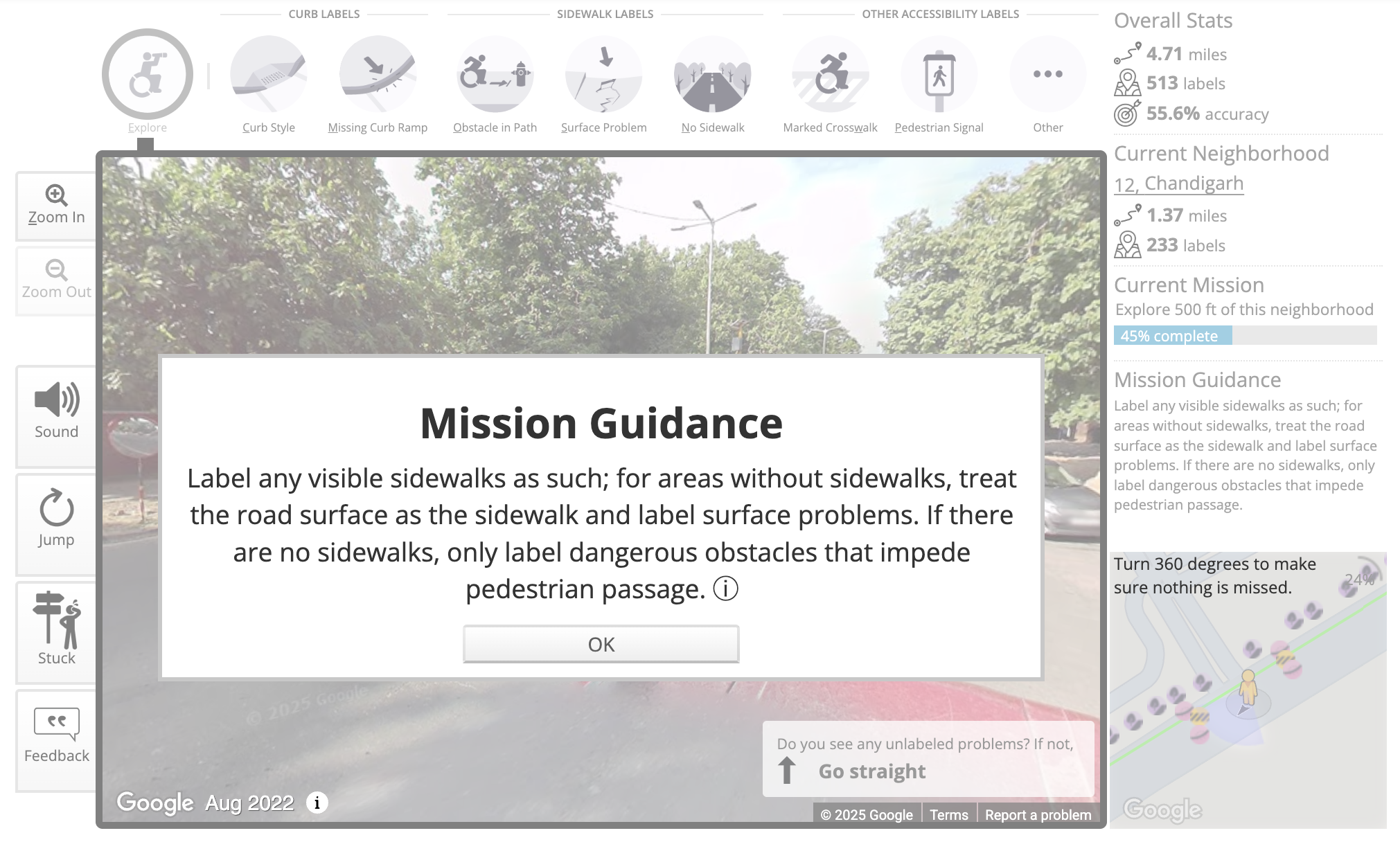}
  \caption{VLM generates segment-specific mission guidance}
  \label{fig:b}
\end{subfigure}
\begin{subfigure}[t]{0.2\textwidth}
  \centering
  \includegraphics[width=\linewidth]{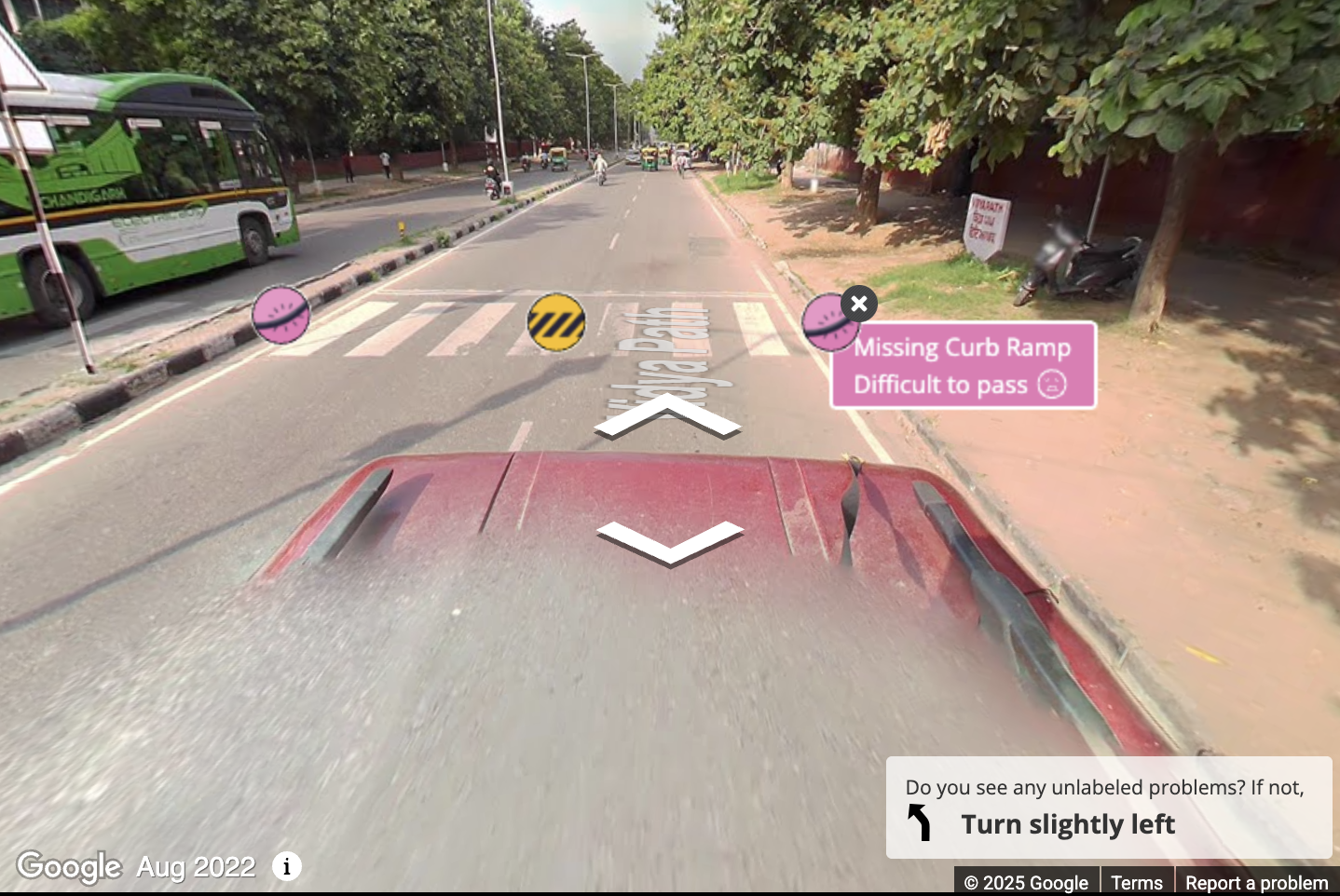}
  \caption{Human annotators label using the adapted Project Sidewalk}
  \label{fig:c}
\end{subfigure}

\caption{\textbf{Systematic overview of our India-adapted Project Sidewalk workflow.}
(a) Population extrapolation from ward-to-sector boundaries for selecting sectors for analysis.
(b) Each street segment in the sector is defined by its OSM road type and start/end Street View panoramas. (c) A visual–language model (VLM) generates mission guidance from these inputs, helping annotators know what accessibility barriers to expect. (d) Annotators label barriers in the adapted Project Sidewalk interface using Indian-specific labels.
}
\label{fig:teaser}
\end{figure*}



\begin{abstract}
Project Sidewalk is a web-based platform that enables crowdsourcing accessibility of sidewalks at city-scale by virtually walking through city streets using Google Street View. 
The tool has been used in 40 cities across the world, including the US, Mexico, Chile, and Europe. In this paper, we describe adaptation efforts to enable deployment in Chandigarh, India, including modifying annotation types, provided examples, and integrating VLM-based mission guidance, which adapts instructions based on a street scene and metadata analysis. Our evaluation with 3 annotators indicates the utility of AI-mission guidance with an average score of 4.66. Using this adapted Project Sidewalk tool, we conduct a Points of Interest (POI)-centric accessibility analysis for three sectors in Chandigarh with very different land uses---residential, commercial and institutional covering about 40 km of sidewalks. Across ~40 km of roads audited in three sectors and around 230 POIs, we identified 1,644 of 2,913 locations where infrastructure improvements could enhance accessibility.

\end{abstract}


\section{Introduction}

The United Nations New Urban Agenda \cite{caprotti2017new} positions equity and inclusion as key principles of modern urban development and transportation. A key problem, however, is developing reliable, cost-effective techniques to collect data on accessibility and make data-driven improvements \cite{froehlich2022future}. Lack of accessibility can take away the joy or even the entire feasibility of a trip for all types of users. Therefore, it is critical to flag accessibility issues for i) informing users where the most accessible areas are, and ii) informing local governments about the required improvements in the public infrastructure. 

Prior work in geolocated accessibility mapping has primarily focused in the US and European contexts \cite{xu2021poilanduse, li2022sidewalkequity}. One of the prominent solutions, Project Sidewalk \cite{ProjectSidewalk} is a web-based platform for mapping and assessing sidewalk accessibility using crowdsourcing and online street imagery. Since a 2016 pilot deployment in Washington DC, Project Sidewalk has grown to 44 cities across 10 countries---over 10k users have contributed 1.4 million labels assessing 26k km of city streets. The tool lets people virtually “walk” through city streets using Google Street View and label accessibility issues such as missing curb ramps, uneven surfaces, or obstacles.
While the gold standard for accessibility audits is manual inspections by experts hired to analyse issues, such methods to flag accessibility issues are laborious and costly \cite{froehlich2022future}.  
However, porting project sidewalk to a new city like Chandigarh which is very different from the US and European context is non-trivial due to the change in local context and different expectations of accessibility in this context across different types of roads - primary, secondary, tertiary roads, etc. The appearance of sidewalks and related features such as curbs, ramps, and surface materials changes significantly for India, which makes the direct use of existing labels confusing for the users (see Figure \ref{fig:interface-composite}). 

\begin{figure}[t]
  \centering

  \begin{subfigure}[t]{0.3\linewidth}
    \centering
    \includegraphics[width=\linewidth]{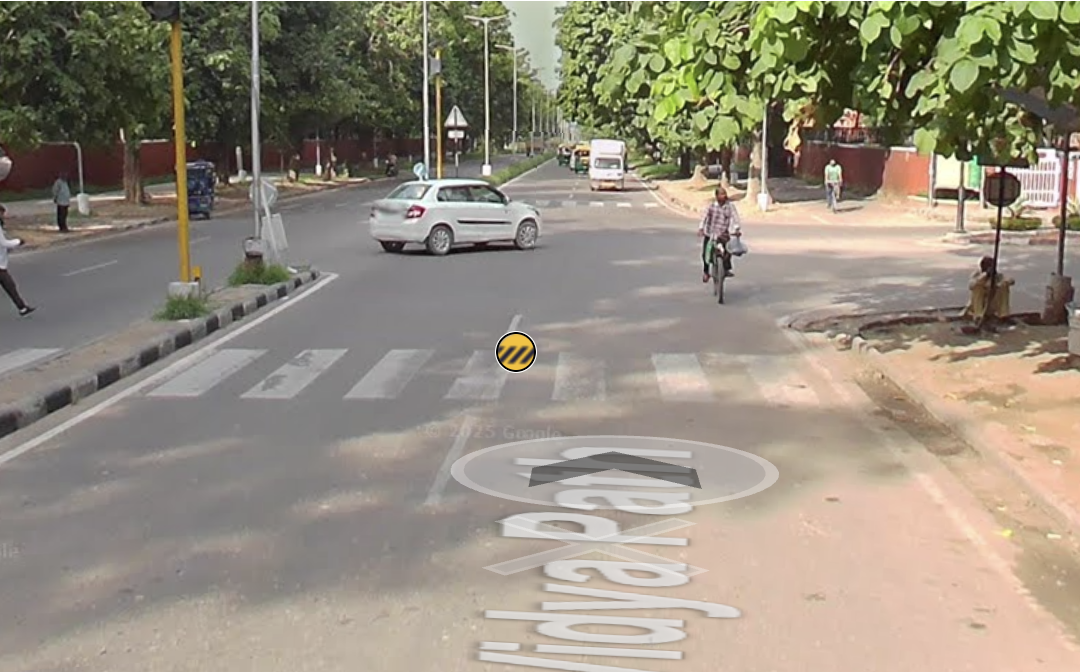}
    \caption{Crosswalk: Paint Fading}
  \end{subfigure}
  \begin{subfigure}[t]{0.3\linewidth}
    \centering
    \includegraphics[width=\linewidth]{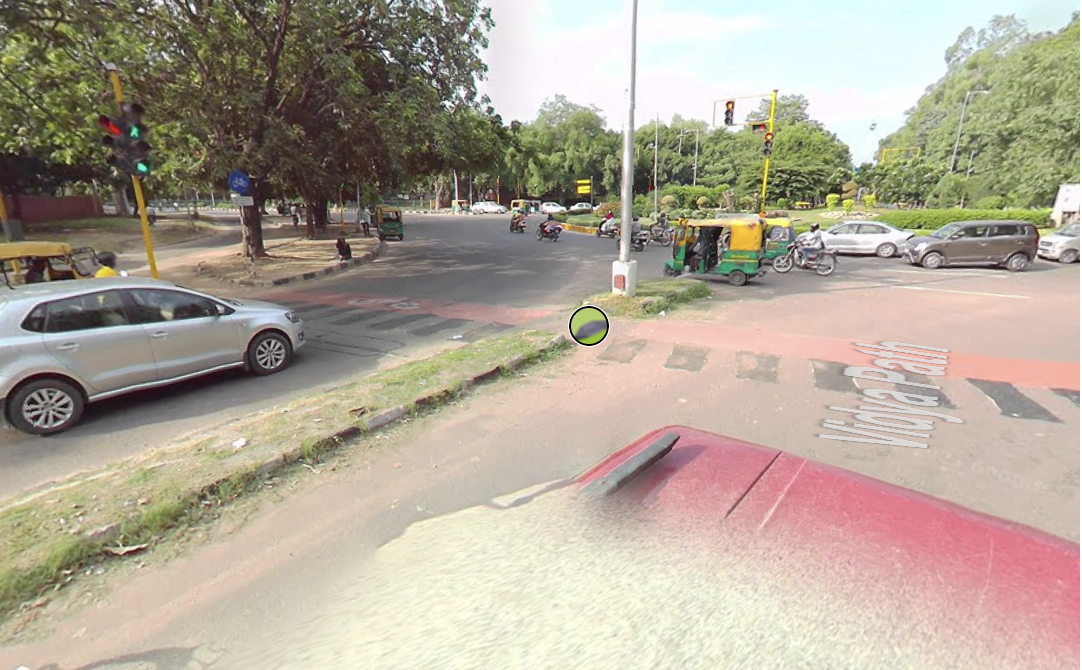}
    \caption{Curb Style: Low Severity}
  \end{subfigure}
  \begin{subfigure}[t]{0.3\linewidth}
    \centering
    \includegraphics[width=\linewidth]{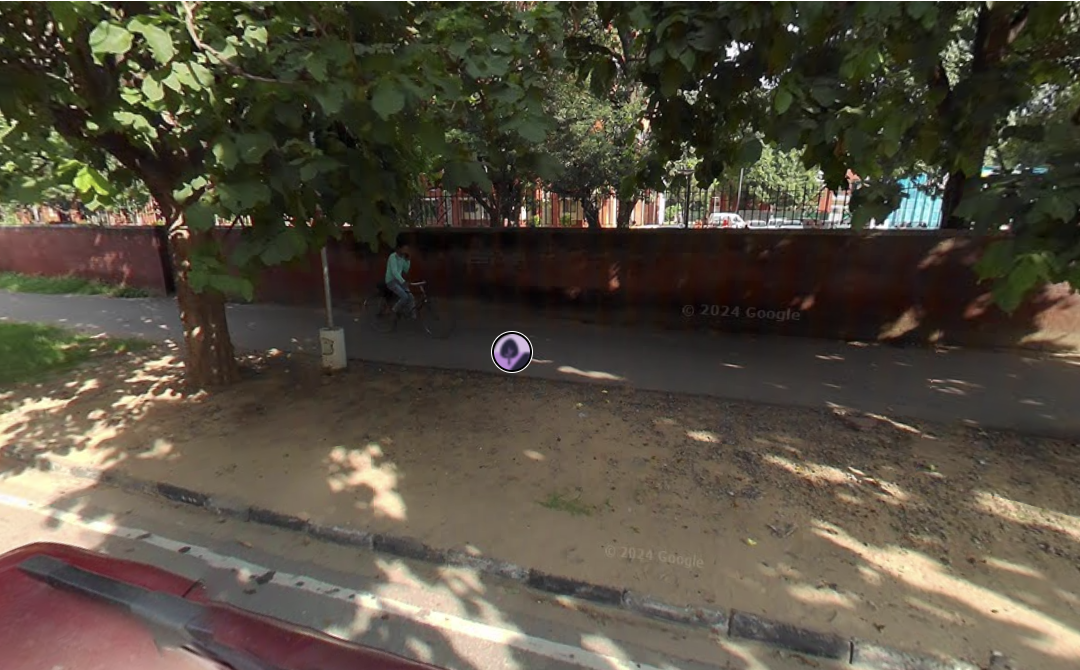}
    \caption{No Sidewalk: Bike Lane}
  \end{subfigure}

  \begin{subfigure}[t]{0.3\linewidth}
    \centering
    \includegraphics[width=\linewidth]{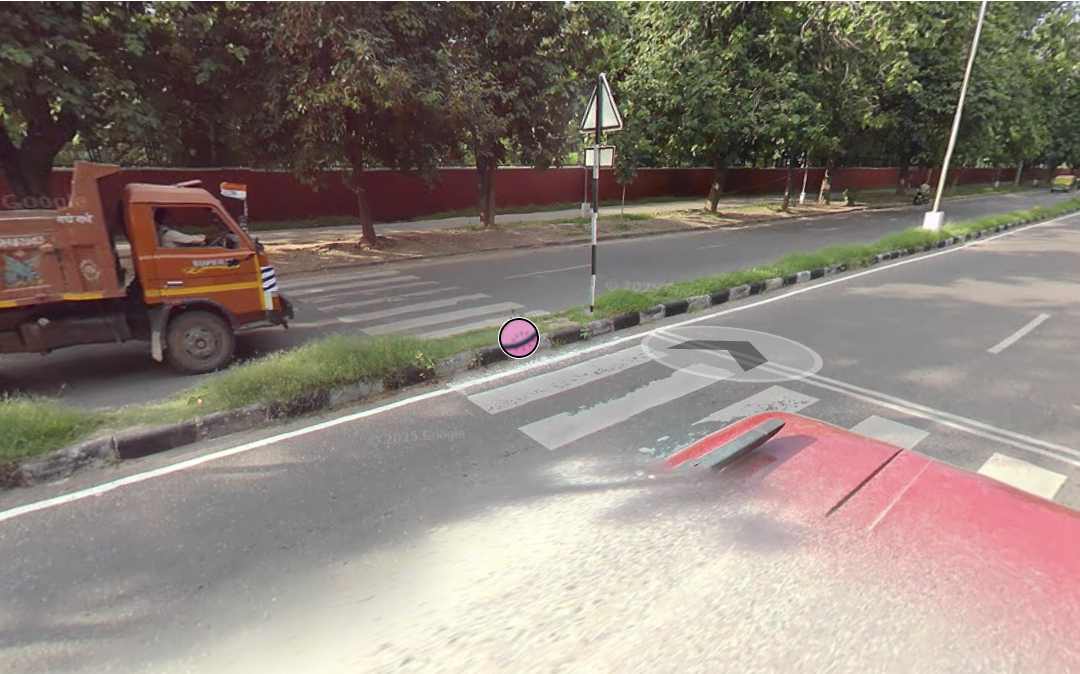}
    \caption{Missing Curb Ramp (Crosswalk)}
  \end{subfigure}
  \begin{subfigure}[t]{0.3\linewidth}
    \centering
    \includegraphics[width=\linewidth]{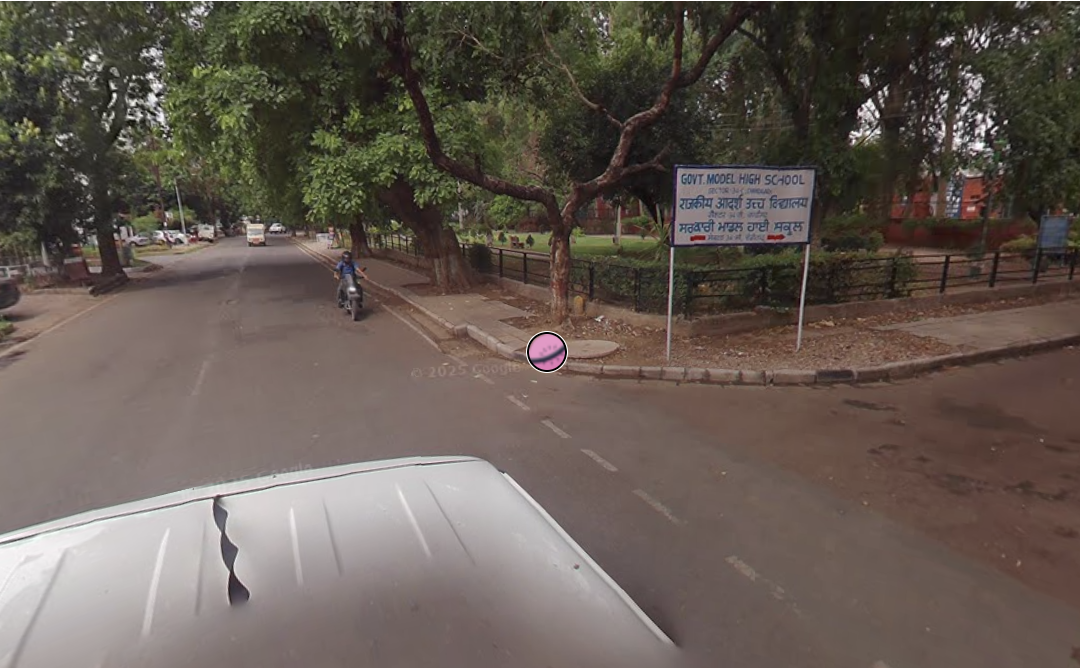}
    \caption{Missing Curb Ramp (Intersection)}
  \end{subfigure}
  \begin{subfigure}[t]{0.3\linewidth}
    \centering
    \includegraphics[width=\linewidth]{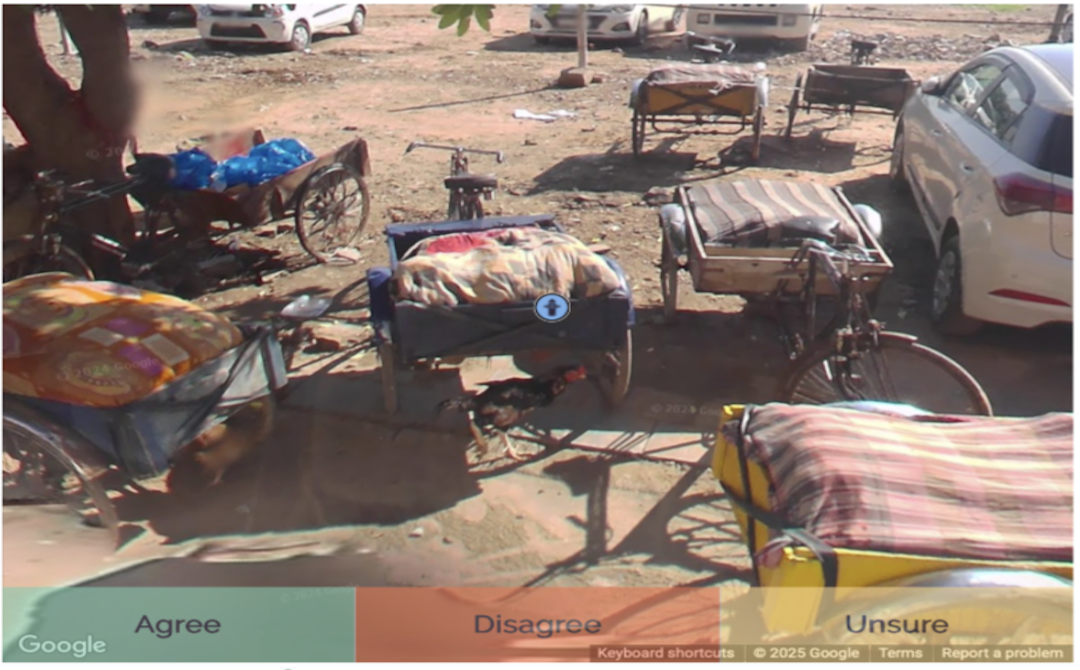}
    \caption{Obstacle: Cart}
  \end{subfigure}

  \begin{subfigure}[t]{0.3\linewidth}
    \centering
    \includegraphics[width=\linewidth]{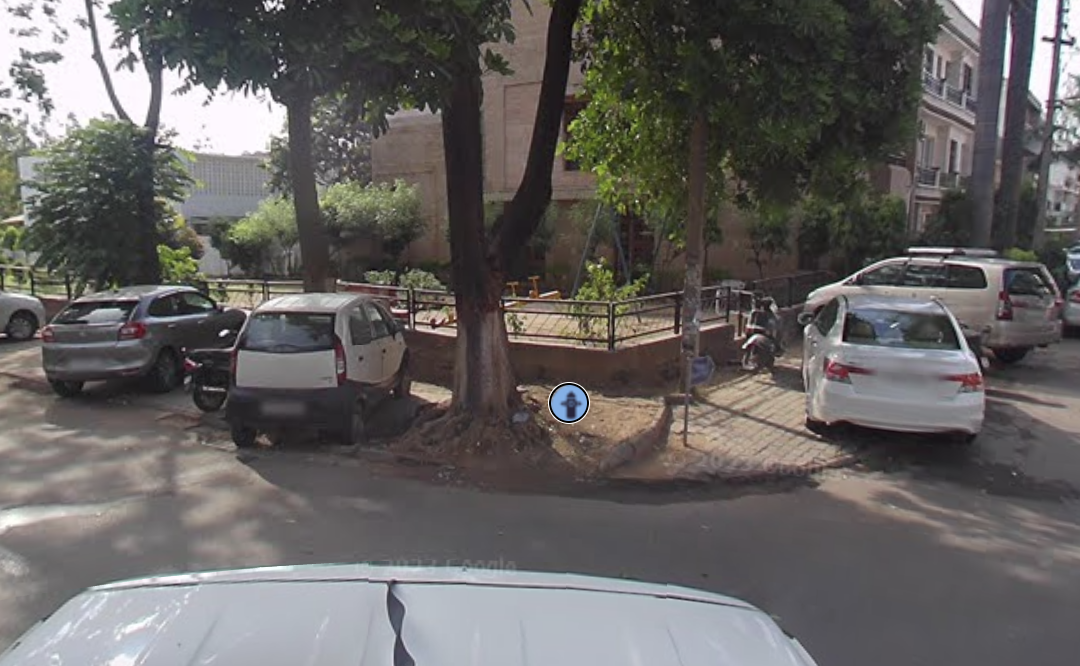}
    \caption{Obstacle: Tree}
  \end{subfigure}
  \begin{subfigure}[t]{0.3\linewidth}
    \centering
    \includegraphics[width=\linewidth]{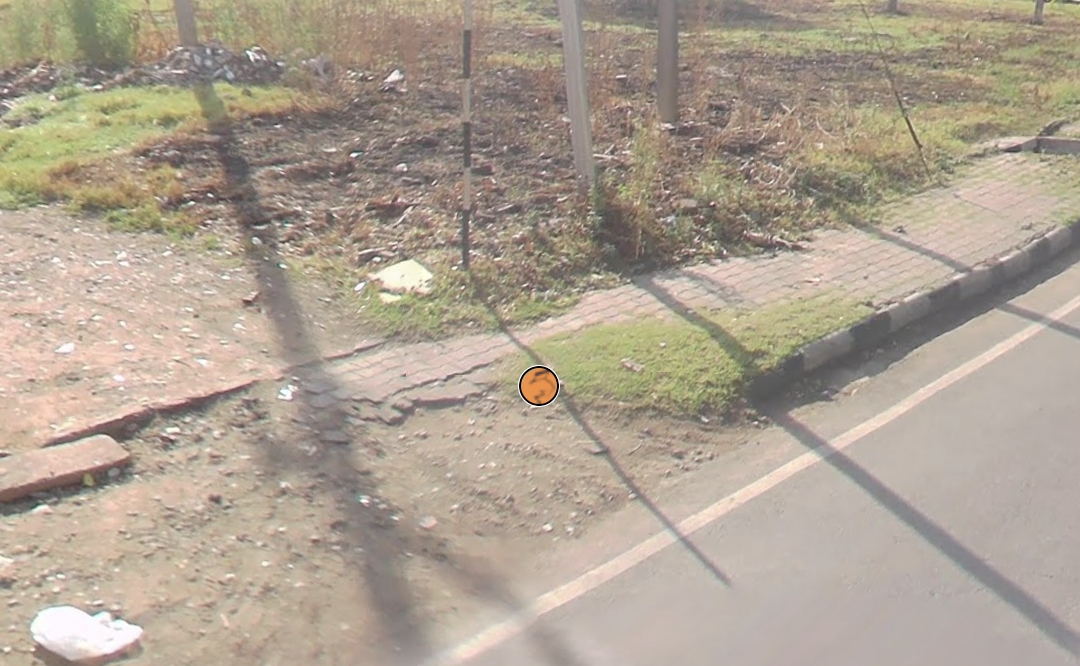}
    \caption{Surface Problem: Gravel}
  \end{subfigure}
  \begin{subfigure}[t]{0.3\linewidth}
    \centering
    \includegraphics[width=\linewidth]{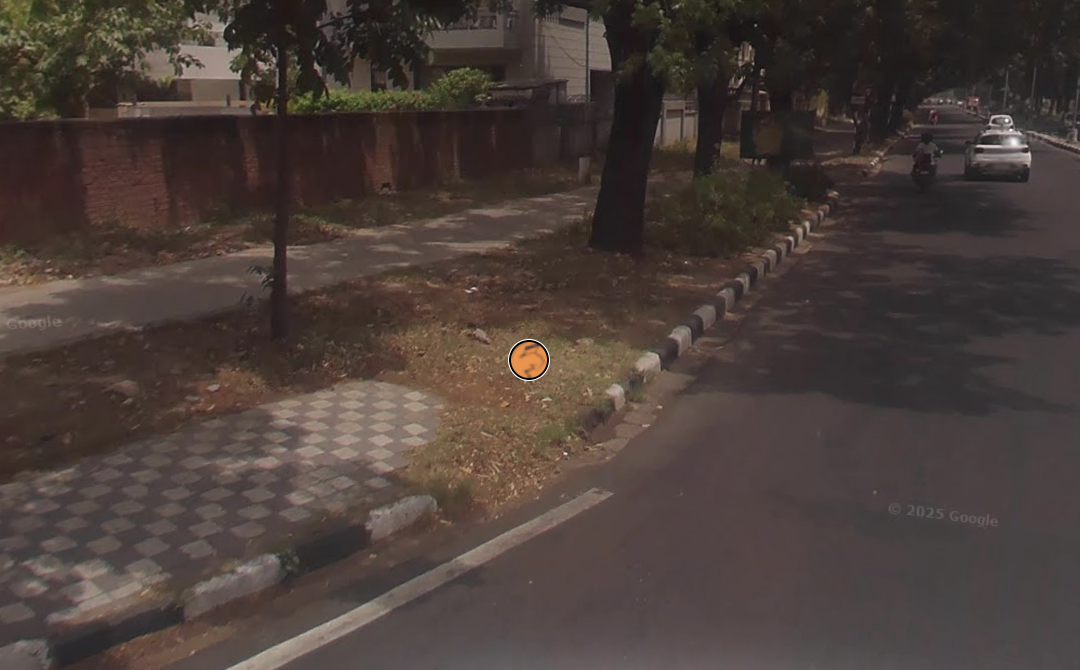}
    \caption{Surface Problem: Mud}
  \end{subfigure}
  \caption{Representative examples of adapted label categories and respective tags in the Indian context. Images were sampled from annotated segments in Chandigarh and illustrate the diversity of visual conditions that motivated the interface redesign.}
  \label{fig:indian_images}
\end{figure}

In this work, we contextualize the Project Sidewalk user experience with locally-enriched tags (Figure \ref{fig:indian_images}) and examples thus enabling the first deployment of Project Sidewalk in India---in a city that is globally recognized for its planning, Chandigarh. 
Apart from this, to provide just-in-time information about annotation strategy on a particular street, we also design and integrate a visual language model (VLM) powered mission guidance (Figure \ref{fig:b}). The mission guidance helps the user understand the context around the road that they are marking and reduces the cognitive overhead required to understand what needs to be marked. To evaluate the VLM-based mission guidance, we evaluate the guidance with three users across metrics for relevance, accuracy and utility.


Given the large population and vast area of the country, an effective prioritization framework is required for identification of accessibility gaps and their remediation by the states. 
Our work is an effort towards highlighting gaps which provide the maximum return on investment through increased accessibility of places in a city. By focusing on accessibility of places instead of all the infrastructure, we provide a first level of prioritization for accessibility in a pilot deployment over three sectors in Chandigarh.
%
%
%
To this end, we focus our study of accessibility in Chandigarh around a walkable one km radius around points of interest in the city including key amenities such as restaurants, transit stops, ATMs, hospitals, schools, etc. (complete list in Table \ref{tab:poi_categories}) across three different land use sectors - Sector 12 (institutional), Sector 34 (commercial) and Sector 45 (residential). We map a total of 230 points of interest (POIs) through open street map and use our adaptations to the project sidewalk for annotating accessibility in the 40 kms of roads surrounding these POIs. Through our detailed annotations of curbs, obstacles, surface problems, crosswalks, etc. we are able to provide a focused set of accessibility gaps for local action. We also create aggregate metrics to provide relative scores for different areas as user insights.
To summarize, our key contributions in the paper include:
\begin{enumerate}
    \item Adaptation of Project Sidewalk to the Indian context through changes in the user interface and an AI-assisted mission guidance per road segment. We evaluate the utility of AI-assistance with three users scoring relevance, accuracy and usefulness and found the mean utility to be 4.66.
    \item Use of this adapted methodology to conduct a cross-sectoral analysis for three sectors in Chandigarh across all points of interest within an institutional, residential and commercial land use. Our analysis is over 200 points of interest in a 40 km region across the three sectors. We observed a long negative tail in segment scores, indicating many segments with substantial problems. Commercial areas have the best overall access, while education and public-service lag. We also found that functional accessibility, e.g., health center accessibility in health-institution center has been given attention but other facilities like transit stops, food joints, etc. are not as accessible in the region.

\end{enumerate}


\section{Related work}
We provide background on the importance of geolocated accessibility maps and situate our work within prior Project Sidewalk deployments and literature in POI accessibility.



\paragraph{The need for geolocated accessibility maps}
The Clean Air Asia study \cite{cleanairasia2011walkability} reviewed pedestrian infrastructure in six Indian cities through walkability surveys, interviews, and policy reviews . It found that sidewalks were often missing or poorly maintained, especially near public transport hubs and schools. A recent study in Kochi used the Continuous Pedestrian Movement (CPM) approach analysis points of disruption in continuous flow of walking from unsafe conditions, obstacles or crossings and showed that existing evaluation methods overlook delays and conflicts in pedestrian movement, especially for vulnerable users \cite{casestudykochi}. A Supreme Court Committee (2023) audit of Delhi’s 1,400 km of Public Works Department (PWD) roads found that 84\% of footpaths failed to meet IRC standards, and only 25\% were usable \cite{newindianexpress2024pwdaudit}. 
In Bengaluru, walkability data from 2023 \cite{opencity2023walkability} showed that pedestrians made up 32\% of all road deaths. Many walkability surveys \cite{globalwalkabilityindex, fuzzyindexdwarka, Bharucha2017AnII} focus on footpath width, cleanliness, or shade, but accessibility for persons with disabilities such as the presence and quality of curb ramps and pedestrian lights is rarely assessed in detail. Thus, while these studies showcase the criticality of the problem, an in-depth localized actionable information is required for implementation by the urban local bodies, which is currently missing.

\paragraph{Adaptation efforts}

Adaptation of accessibility mapping to\-ols such as Project Sidewalk to new geographies requires contextual adjustments in both interface design and labeling strategy. Prior adaptations in cities across the United States, Mexico, and the Netherlands demonstrated that while the underlying platform for virtual sidewalk auditing remains robust, cross-country deployments demand localization of examples, severity ratings, and surface-type taxonomies to match local infrastructure characteristics and cultural contexts \cite{ProjectSidewalk, mexico-sidewalk}. For instance, the Mexico City adaptation required retraining annotators using street-level imagery reflecting informal sidewalks and heterogeneous curb designs \cite{mexico-sidewalk}.

Specifically, the use of LLMs is also becoming prevalent in annotation use cases. They are commonly used as auto labelers (LLM assigns labels directly), as assistants that give guidance or hints to human annotators, and as part of label aggregation where LLM outputs are combined with crowd labels to improve quality \cite{tan2024llm4annotation, li2024crowdllm}. The closest work uses LLM based annotation guidance for NLP tasks while taking inter-annotator agreement from 0.593 to 0.84 \cite{bibal2025automating}. Since our task still requires the human annotator to parse through 3D space using street views and create annotations, measurement of label agreement directly is harder. However, we use a likert scale for users to annotate the perceived utility of guidance in different scenarios. 

\paragraph{Studying place accessibility}
Recent studies increasingly employ \textit{POI-centric audits} to evaluate walkable access to essential urban services such as schools, hospitals, and transit stops. Using circular or network-based walkability buffers—typically 400 m to 1 km—they estimate how physical conditions and sidewalk continuity shape pedestrian access (Wang, Chen, and Liu 2018; Xu et al. 2021; OpenCity 2023). These spatially explicit approaches reveal micro-scale inequities that aggregate citywide indicators often overlook. Building on this framework, we assess accessibility within 1 km walkable buffers around representative POIs in Chandigarh, linking sidewalk-level annotations to sector-level accessibility metrics. We also ensure that there is sufficient GSV coverage around the POIs before selecting the sector.

\section{The context difference}

\subsection{The location}
Chandigarh is one of the earliest planned cities in India.
The city was originally planned for about five lakh people and covered an area of around 70 square kilometers, which has now expanded to about 113 sq km \cite{chandigarh2019landuse}. Its layout follows a sector-based grid, with sectors numbered from 1 to 56, and was meant to be a self-contained city surrounded by a rural belt to control future growth. 
The residential sectors were planned as self-sufficient neighborhoods, each with schools, shops, community centers, and parks. This clear and organized planning makes Chandigarh an ideal case for studying accessibility and walkability. 
The road network of Chandigarh is designed in a grid-iron pattern and is based on seven types of roads, called the 7 Vs (Les Sept Voies) \cite{patle2021sevenvs, chandigarh2019traffic}. These roads create a clear hierarchy for both vehicles and pedestrians from fast traffic carrying roads in V1 to pedestrian paths and cycling tracks in V7 running through carefully planned green areas. Our observation has been that reachability of POIs depends on a mixture of these road types and therefore, accessibility focus is required across road types.

\subsection{The law}
In India, the legal framework mandates provisions for accessibility through the Rights of Persons with Disability Act (RPwD) 2016 \cite{rpwd}.
Accessibility in the physical environment has a trickle down effect for all citizens including those with disabilities. 
The Harmonised Guidelines for Barrier-Free Built Environment \cite{mohua2016harmonisedguidelines, niua2021harmonisedprelude} combine legacy manuals into one practical guide for universal design. These guidelines give clear standards for planners and local bodies to create spaces that are accessible to persons with disabilities, the elderly, and others with limited mobility, promoting more inclusive cities across India.

\subsection{Sector Selection}

Chandigarh’s Master Plan zones sectors as residential, commercial, institutional, or mixed. The city has three phases-Phase I (1-30), Phase II (31-46), Phase III (48-56, 61, 63)-plus nearby areas like Khuda Alisher, Manimajra, and Rajindra Park. Almost 39 sectors in Phases I–II are residential; key institutional hubs include Sectors 1, 12, 14, 17 (Capitol Complex, PGI, Panjab University). Major commercial areas cluster in Sectors 26, 34, 43 (City Centre and Sub-City Centres). Phase III is mainly residential with mixed-use corridors along Vikas Marg.

To select representative sectors for our study, we used both land-use classification and population distribution. We began with ward-wise population data and mapped it to the sector level by assuming that population is uniformly distributed within each ward. 
The sector with the highest residential population was found to be Sector 45, which falls in Phase II. This result aligns with the Master Plan, which identifies Phase II as a high-density zone planned to accommodate a larger residential population.

For the institutional category, we compared three major institutional sectors: Sector 12 (hospital area), Sector 1 (court complex), and Sector 14 (university campus). Among these, the hospital sector (Sector 12) was selected as the representative institutional area, since the institution Postgraduate Institute of Medical Education and Research (PGIMER) attracts large numbers from resident as well as traveling population. The sector spans approximately 277 acres (1.12 sq km) and contains multiple hospital buildings for different medicine specializations.

\begin{table*}[t]
\centering
\small

\begin{tabularx}{\textwidth}{@{}l X@{}}
\toprule
\textbf{Category} & \textbf{Included POI types} \\
\midrule
Financial services &
\texttt{bank}, \texttt{atm}, \texttt{finance}, \texttt{accounting} \\
Education &
\texttt{primary\_school}, \texttt{school}, \texttt{secondary\_school}, \texttt{university} \\
Healthcare &
\texttt{doctor}, \texttt{hospital}, \texttt{pharmacy}, \texttt{health}, \texttt{dentist}, \texttt{drugstore} \\
Public service &
\texttt{local\_government\_office}, \texttt{political} \\
Transport &
\texttt{parking}, \texttt{car\_rental}, \texttt{car\_repair} \\
Food &
\texttt{cafe}, \texttt{food}, \texttt{bar}, \texttt{bakery}, \texttt{restaurant}, \texttt{grocery\_or\_supermarket}, \texttt{meal\_takeaway} \\
Religious &
\texttt{place\_of\_worship}, \texttt{hindu\_temple} \\
Utilities &
\texttt{gas\_station} \\
Commercial &
\texttt{store}, \texttt{beauty\_salon}, \texttt{clothing\_store}, \texttt{electronics\_store}, \texttt{florist}, \texttt{furniture\_store}, \texttt{general\_contractor}, \texttt{gym}, \texttt{hardware care}, \texttt{real\_estate\_agency}, \texttt{hardware\_store}, \texttt{travel\_agency}, \texttt{storage}, \texttt{lawyer}, \texttt{lodging}, \texttt{moving\_company}, \texttt{home\_goods\_store} \\
Social &
\texttt{park}, \texttt{movie\_theatre} \\
\bottomrule
\end{tabularx}
\caption{POI categories and included types.}
\label{tab:poi_categories}
\end{table*}

We used Foursquare Places data \cite{foursquare2024categories} to identify commercial areas. From its 1,000+ POI categories, we selected those common in India---restaurants, retail shops, gyms, cafés, malls, beauty salons, repair services, and entertainment spaces and counted them in Sectors 26, 34, and 43. The totals were 162 (Sector 26), 171 (Sector 34), and 164 (Sector 43). Since Sector 34 had the most commercial POIs, we chose it as the representative commercial sector. These three sectors---45 (residential), 34 (commercial), and 12 (institutional) cover about 40 km of sidewalks (after POI filtering; see next section). This set lets us compare pedestrian accessibility across land-use types around POIs using the Project Sidewalk tool.

\subsection{Points of Interest selection}
For extracting Points of Interest (POIs) over all sectors, we divided each sector into small spatial segments and sampled one point from each segment. For every sampled point, we used the Google Places API to collect all POIs within a 400-meter radius. This process initially gave us 23,136 POI entries. Since nearby points often returned the same POIs, we removed duplicates based on their latitude and longitude values. After cleaning, we obtained 10,128 unique POIs. From the final dataset, we identified 98 distinct POI types across the selected sectors. We grouped them into 10 categories using OSM-style tags; the category definitions and example subtypes are shown in Table~\ref{tab:poi_categories}.

To extract walking paths around each POI, we first explored the local street network using a road graph built with OSMnx. Each POI served as a starting point, and paths were traced in all directions up to 1 km using a depth-first search (DFS) over the network edges. 
Next, we assessed Google Street View (GSV) coverage around these paths to ensure visual data availability for annotation. For every POI, we built a 1 km buffer and divided it into small square cells (about 60–80 m). The centroid of each cell was used to query the nearest available GSV location. Cells with valid coverage were marked as accessible, and their geometries were merged to form a coverage area. 
We created a GSV coverage map using the Street view API \cite{streetviewapi}
We then retained only the path segments where at least 75\% of the geometry overlapped with Street View coverage. These filtered paths formed the final set used for accessibility labeling and analysis.

\begin{figure*}[ht]
  \centering
  \begin{subfigure}[b]{0.48\textwidth}
    \centering
    \includegraphics[width=\linewidth]{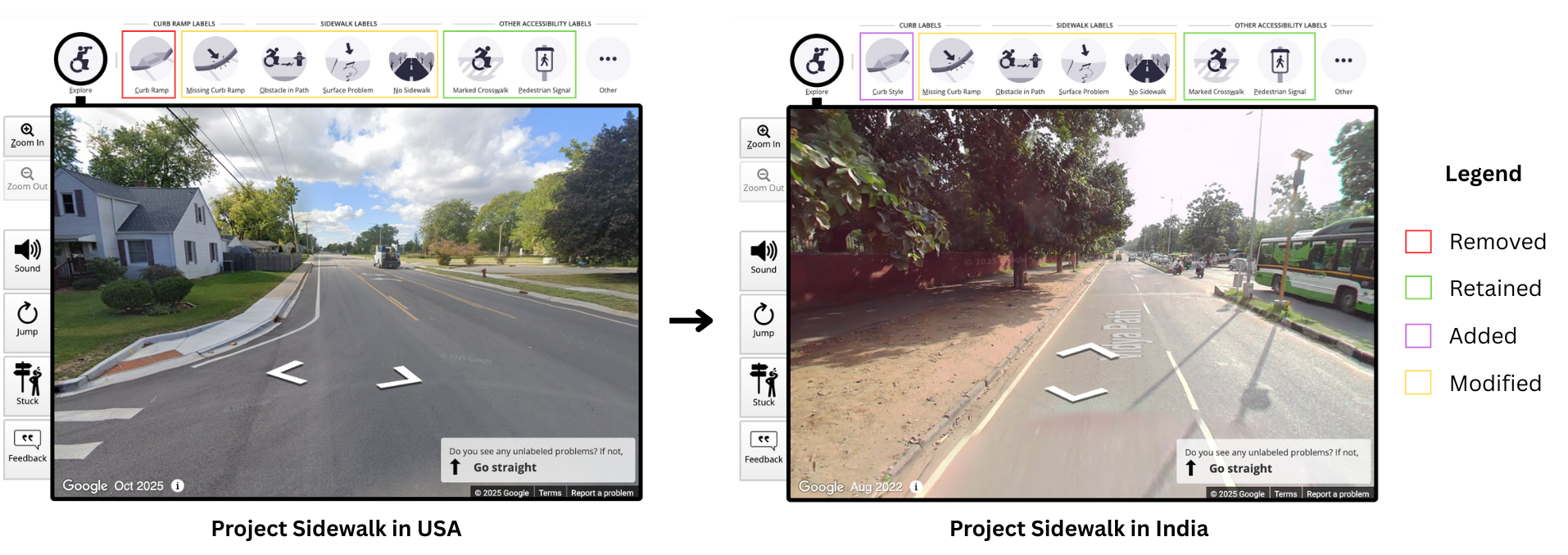}
    \caption{Mapping of original U.S. labels to the adapted India label set, showing retained, modified, added, and removed categories.}
    \label{fig:single-row}
  \end{subfigure}
  %
  \begin{subfigure}[b]{0.48\textwidth}
    \centering
    \begin{minipage}[t]{0.2\linewidth}
      \centering
      \includegraphics[width=\linewidth]{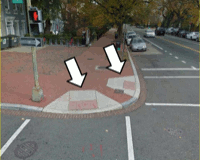}
      {\footnotesize US: Low\par}
    \end{minipage}
    \begin{minipage}[t]{0.2\linewidth}
      \centering
      \includegraphics[width=\linewidth]{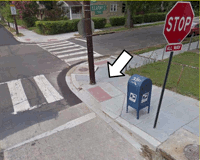}
      {\footnotesize US: Medium\par}
    \end{minipage}
    \begin{minipage}[t]{0.2\linewidth}
      \centering
      \includegraphics[width=\linewidth]{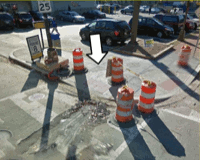}
      {\footnotesize US: High\par}
    \end{minipage}
    
    \begin{minipage}[t]{0.2\linewidth}
      \centering
      \includegraphics[width=\linewidth]{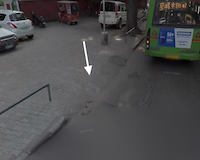}
      {\footnotesize India: Low\par}
    \end{minipage}
    \begin{minipage}[t]{0.2\linewidth}
      \centering
      \includegraphics[width=\linewidth]{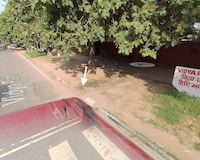}
      {\footnotesize India: Medium\par}
    \end{minipage}
    \begin{minipage}[t]{0.2\linewidth}
      \centering
      \includegraphics[width=\linewidth]{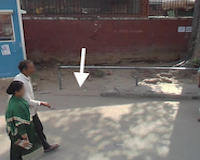}
      {\footnotesize India: High\par}
    \end{minipage}
    \caption{Low/medium/high severity examples for curb transitions in the U.S. (top) and India (bottom).}
    \label{fig:grid-row}
  \end{subfigure}

  \caption{Adaptations to the Project Sidewalk interface for deployment in India. (a) Overview of label-set changes, showing which categories were retained, removed, renamed, or newly introduced to reflect Indian pedestrian infrastructure. (b) Revised severity examples illustrating how the original U.S. Curb Ramp label maps to the broader Indian Curb Style category, while preserving the three-level severity scale.}
  \label{fig:interface-composite}
\end{figure*}

\section{Adapting Project Sidewalk to India}

Sidewalk accessibility in Indian cities differs not only in quality but in form. Pedestrians frequently walk on shared carriageways, informal shoulders, or discontinuous footpaths rather than standardized sidewalks; curb ramps are rare, surface transitions are improvised, and vendor stalls, drainage channels, and parked two-wheelers regularly occupy pedestrian space on sidewalks. These conditions make several accessibility barriers highly visible in India that may not be present in prior Project Sidewalk deployments, while elements such as ADA-style curb ramps or consistent pedestrian signals—which are central to the original interface—appear infrequently or in heterogeneous forms.

These characteristics of the Indian context raised two key adaptation needs:
\begin{itemize}
    \item \textbf{Label mismatch:} Several default labels, tags, and example images did not clearly apply to the types of pedestrian barriers visible in Chandigarh’s street imagery.
    \item \textbf{Ambiguity in what to annotate:} New annotators struggled to decide which labels were appropriate on streets without formal sidewalks or where walking space was visually ambiguous.
\end{itemize}

To address these issues, we introduced two corresponding adaptations. First, we modified the interface by updating label names, tags, and example images to better reflect Indian streetscape conditions—for instance, replacing the ``Curb Ramp'' label with a broader ``Curb Style'' category that captures sloped, stepped, or improvised curb transitions. Second, we integrated a visual language model (VLM)-assisted mission guidance system with Project Sidewalk that provides brief, context-aware prompts at the start of each street segment, helping annotators understand which accessibility features are likely to appear and what to focus on.

These changes retain the core crowdsourcing workflow of Project Sidewalk while making the tool more aligned with the infrastructural and visual characteristics of Indian cities.

\subsection{Interface Redesign}
We describe each adaptation below, beginning with changes to the user-facing interface, followed by the integration of VLM-based guidance.
To align the platform with the types of pedestrian barriers commonly encountered in Indian streetscapes, we adapted the Project Sidewalk interface through targeted changes to the labeling schema, tag vocabulary, and example imagery. 
Our modifications focus on (i) renaming or restructuring labels to better capture the range of infrastructure conditions observed in the Indian context, and (ii) localizing the tag set and visual examples shown to annotators.

Curb ramps in India vary significantly in style, quality, and construction---ranging from formal concrete ramps to improvised slopes, stepped transitions, or partial modifications. The original label, which focused on the presence or absence of a standardized ramp design, did not adequately capture this diversity. Thus, the \textit{Curb Ramp} label was adapted to \textit{Curb Style} which allows annotators to describe how the pedestrian path transitions to the road surface, including sloped edges, stepped drops, broken curbs, or drainage gaps. This reframing maintains the intent of the original label---assessing whether curb transitions are navigable---but does so in a way that accommodates the variability observed in local infrastructure.
As illustrated in Figure~\ref{fig:interface-composite}, the replacement of the original \emph{Curb Ramp} label with the India-specific \emph{Curb Style} label is paired with updated low/medium/high severity examples drawn from Chandigarh, ensuring that the interface preserves the core annotation workflow of Project Sidewalk while aligning visual guidance with the local streetscape conditions.

\begin{table*}[t]
\small
\centering
\begin{tabularx}{\textwidth}{@{}p{0.16\textwidth}X X X@{}}
\toprule
\textbf{Label} & \textbf{Retained Tags} & \textbf{New Tags} & \textbf{\textcolor{gray}{Removed Tags}} \\ 
\midrule

\textbf{Curb Ramp (US)} 
& --- 
& --- 
& \textcolor{gray}{Narrow, steep, points into traffic, not enough landing space, missing tactile warning, pooled water} \\

\textbf{Curb Style (India)} 
& --- 
& not level with street, pooled water, not visible, debris
& --- \\

\textbf{Missing Curb Ramp} 
& unclear if needed, alternate route present, no alternate route 
& --- 
& --- \\

\textbf{Obstacle in Path} 
& tree, pole, vegetation, parked cycle, construction, sign, stairs 
& parked car, carts, drainage, electric box 
& \textcolor{gray}{narrow, parked bike, garbage, fire hydrant, mailbox, recycle bin} \\

\textbf{Surface Problem} 
& bumpy, cracks, grass, narrow, uneven/slanted, height difference, cobblestone, sand/gravel, broken
& sand/gravel/mud 
& \textcolor{gray}{brick, utility panel, debris, rail} \\

\textbf{No Sidewalk} 
& ends abruptly, street has no sidewalk, pedestrian lane marking, gravel/dirt road
& too dirty/cluttered 
& \textcolor{gray}{shared pedestrian/car space, street has a sidewalk} \\

\textbf{Crosswalk} 
& paint fading, broken surface, uneven surface, bumpy, very long crossing, brick/cobblestone 
& --- 
& \textcolor{gray}{rail/tram track} \\

\textbf{Pedestrian Signal} 
& hard to reach buttons, one button, two buttons, tactile audible buttons & --- & --- \\

\bottomrule
\end{tabularx}
\caption{Label and tag adaptations made to Project Sidewalk for deployment in Chandigarh, India. 
\textbf{Curb Ramp} is removed and replaced with \textbf{Curb Style} to support the heterogeneous curb transitions found in Indian streetscapes. Retained tags are shown in the second column, newly added India-specific tags in the third, and tags removed from the U.S. version in \textcolor{gray}{gray}.}
\label{tab:label_comparison_gray}
\end{table*}

Other labels were retained but updated through additions or removals of tags. For example, the \textit{Obstacle in Path} label now includes India-specific obstructions such as street vendors, drainage channels, parked two-wheelers, and temporary construction debris, while tags less relevant to the Indian setting (e.g., fire hydrants, mailboxes) were removed. The \textit{Surface Problem} label was expanded to include conditions such as loose brick, mud, and unpaved shoulders. The \textit{No Sidewalk} label was also revised to distinguish between streets with no available pedestrian space and those where space exists but is unusable due to clutter or encroachment. Table~\ref{tab:label_comparison_gray} summarizes these interface changes, showing which tags were retained, added, or removed for each label.

All example images shown in hover tooltips were replaced with images from streets in Chandigarh to reduce annotation ambiguity and make severity judgments more intuitive. These updates preserve the overall structure of the Project Sidewalk interface while ensuring that the taxonomy, visual examples, and annotation expectations reflect conditions commonly seen in Indian streetscapes.

\subsection{VLM-Assisted Mission Guidance}

To reduce the cognitive load on annotators and improve the consistency of labeling in a new geographic context, we introduce an AI-driven mission guidance feature that provides brief, context-aware instructions at the start of each street segment. Annotators in India encounter highly variable pedestrian environments—ranging from formal sidewalks to shared pedestrian–vehicle corridors—making it difficult for new users to infer which labels are relevant without additional support. The goal of the guidance is not to automate labeling, but to supply just-in-time cues that help annotators decide what to look for before beginning a task.

We used \textbf{Gemini's} \texttt{gemini-2.5-flash} model to generate guidance, which is triggered under three conditions: (1) when a mission begins, (2) when the annotator moves to a new street segment, and (3) when the user selects the \texttt{Jump} option to relocate to a different part of the city. Each guidance message is produced from two inputs: the OpenStreetMap (\texttt{OSM}) road type for the current segment (e.g., \textit{residential}, \textit{secondary}, \textit{tertiary}) and the first and last Google Street View (GSV) panoramas associated with that segment. These panoramas are automatically retrieved, encoded, and passed to the model alongside a structured prompt that instructs the VLM to generate practical, India-aware annotation advice.

The use of road type is intentional: in Indian cities, residential roads often lack dedicated sidewalks, whereas primary and arterial roads are more likely to have raised walking space, curb infrastructure, or marked crossings. The guidance therefore adapts accordingly—for example, prompting annotators on residential streets to look for obstacles and surface problems on the road surface itself, while prompting curb-related or crosswalk-related checks on higher-order roads. Similarly, supplying both start and end panoramas reduces misinterpretation caused by abrupt changes in sidewalk presence, which are common within short distances in India.

The output is a short, natural language instruction displayed in a popup and in a persistent status panel above the minimap as shown in Figure \ref{fig:b}. It does not create labels, but instead orients annotators toward the most relevant label categories for that segment, based on the India-adapted labeling interface described in the previous section.

\begin{table*}[t]
\centering
\small
\setlength{\tabcolsep}{6pt}
\renewcommand{\arraystretch}{1.18}
\begin{tabular}{C{0.23\textwidth} C{0.12\textwidth} L{0.57\textwidth}}
\toprule
\textbf{Panorama} & \textbf{Road Type} & \textbf{VLM Guidance Message} \\
\midrule

\includegraphics[width=0.8\linewidth,height=28mm,keepaspectratio]{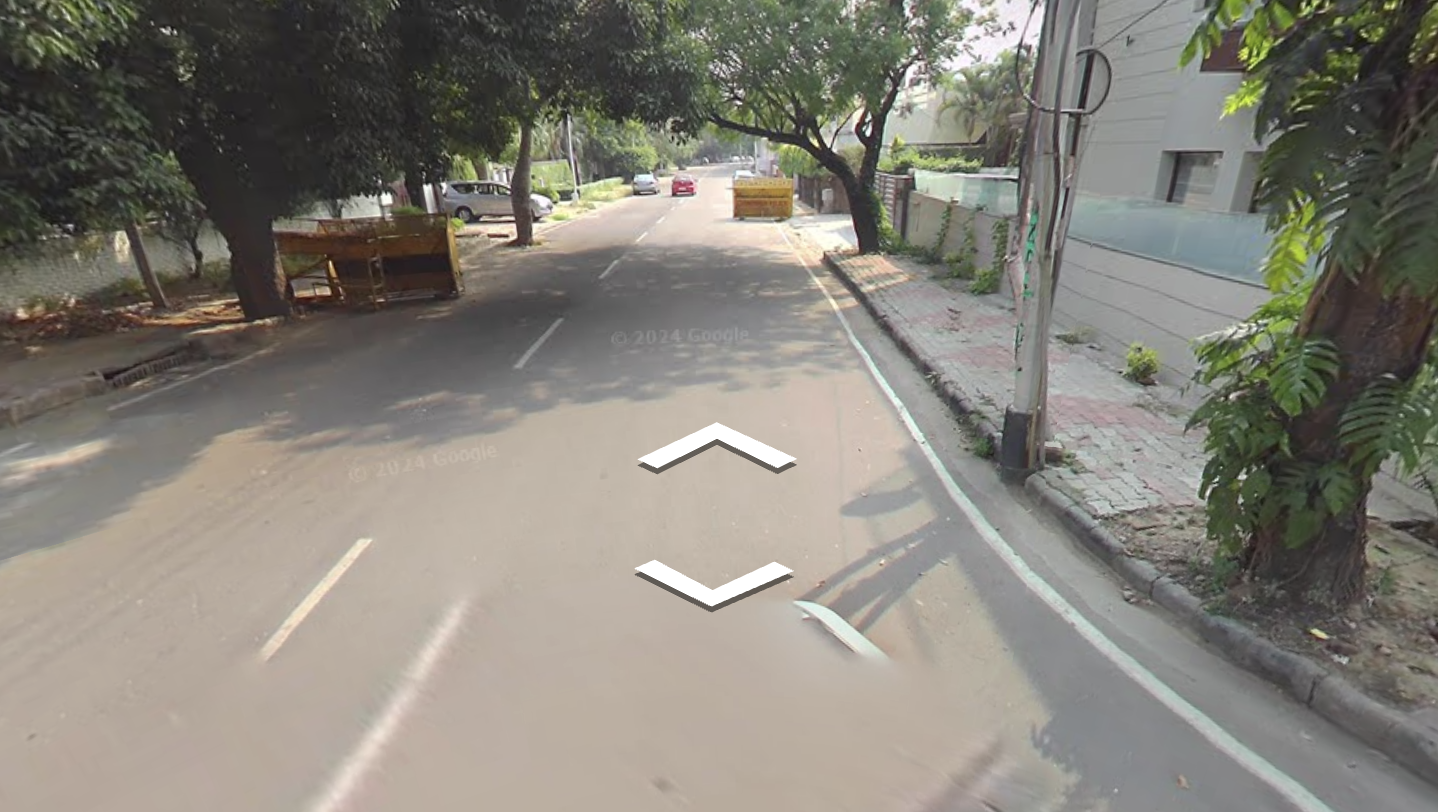} &
\textbf{Tertiary} &
Since this is a tertiary road, expect a constructed sidewalk on at least one side. Focus on the paved sidewalk (right side) for broken pavers, uneven tiles, or tree-root damage and mark surface problems where the bricks are lifted or missing. Watch for obstacles like the utility pole and construction barricades blocking the walkway. Only check for curb ramps or missing curb ramps at visible crosswalks or intersections further along the segment. \\

\addlinespace[2pt]

\includegraphics[width=0.8\linewidth,height=28mm,keepaspectratio]{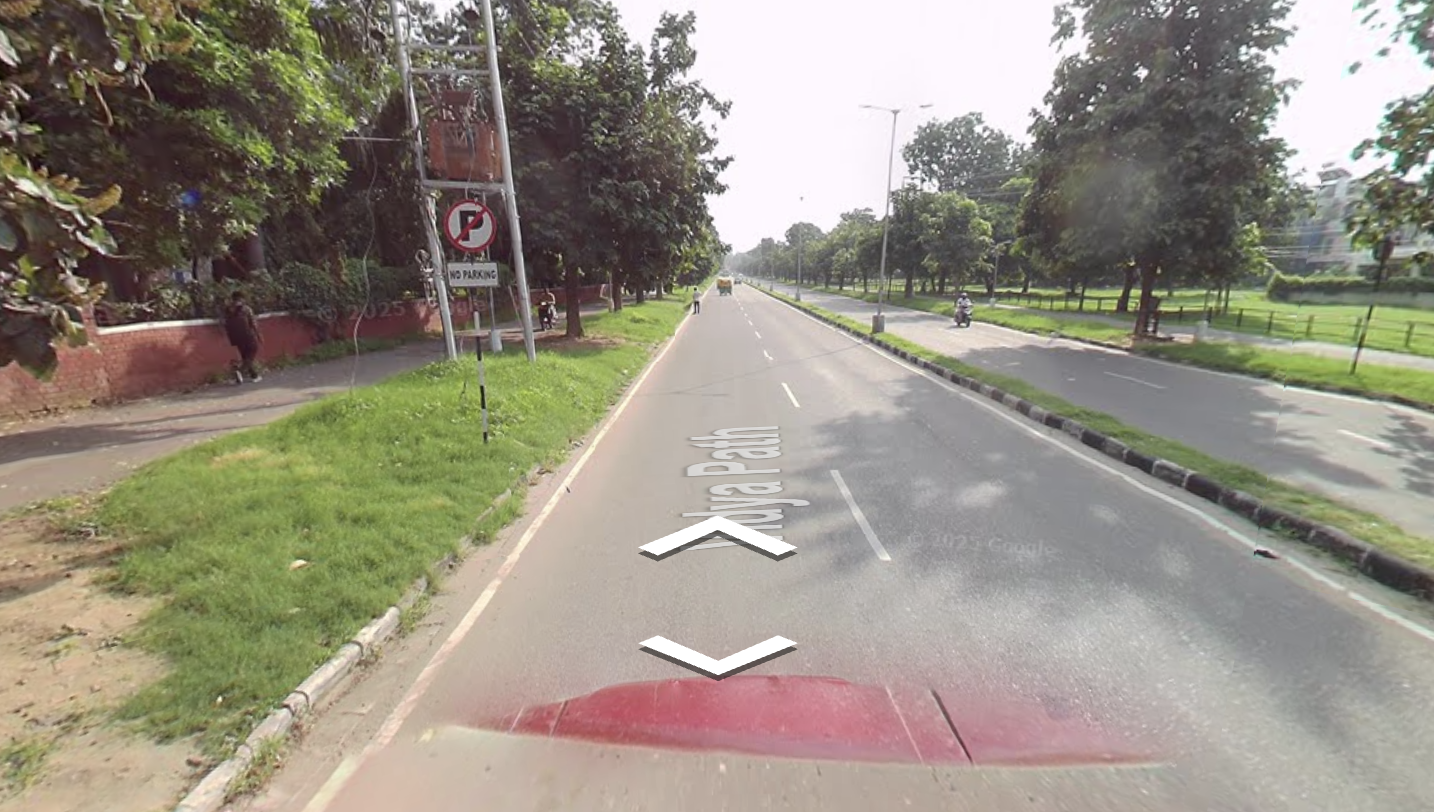} &
\textbf{Secondary} &
This is a secondary road, so expect a proper sidewalk along the edges rather than using the road as the pedestrian path. Scan both sides for a constructed walkway vs.\ grass verge to decide if ``No Sidewalk'' applies. Prioritize checking for missing curb ramps at crosswalks, intersections, and where the paved path meets the road, and note any obstacles like poles or signage blocking the walking space. \\

\addlinespace[2pt]

\includegraphics[width=0.8\linewidth,height=28mm,keepaspectratio]{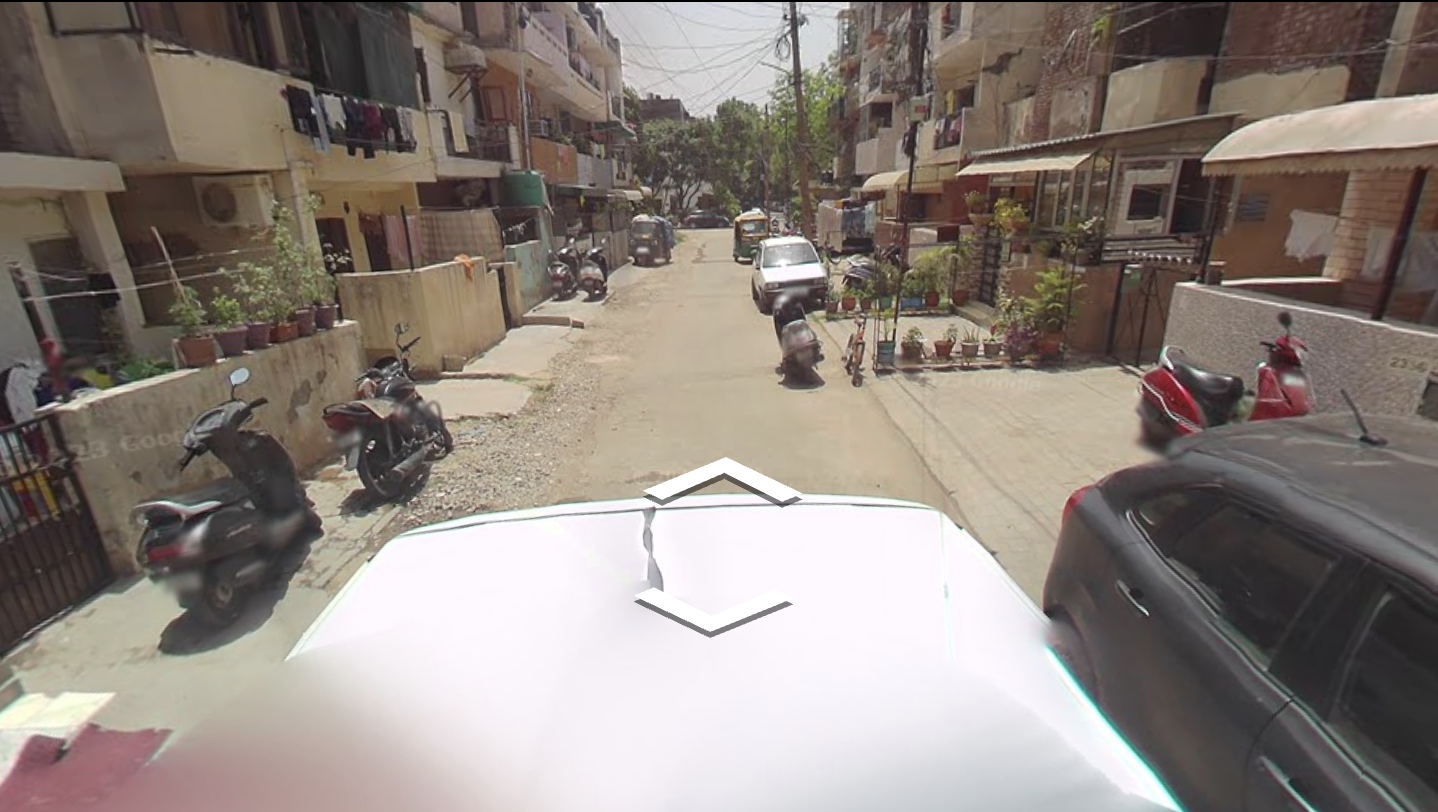} &
\textbf{Residential} &
This is a narrow residential street, so treat the road itself as the pedestrian path. Focus on obstacles like parked scooters, plants, and uneven road edges, and note any surface problems where the pavement is broken or gravel replaces asphalt. Only check for curb ramps at visible crosswalks or intersections, but otherwise prioritize obstacle and surface labeling over curb-related tags. \\

\bottomrule
\end{tabular}
\caption{Examples of VLM-generated guidance messages paired with roadway contexts. Each message instructs annotators on what to focus on based on the road type visible in the panorama.}
\label{tab:guidance_images}
\end{table*}

Table~\ref{tab:guidance_images} shows sample guidance messages alongside the corresponding input panoramas used to generate them. To assess the quality of these messages, we evaluate each one along three dimensions: Relevance (how well the guidance reflects what is visible in the images), Usefulness (the extent to which the message provides actionable information for annotation), and Accuracy (whether the guidance is factually correct given the street context). Each dimension is rated on a 5-point Likert scale. 


\section{Accessibility Metrics}
We adapt the approach of the AccessScore metric from prior work on sidewalk accessibility mapping \cite{li2022sidewalkequity} but simplify it for our context. Our goal is to quantify how accessible a sidewalk segment is based on the presence and severity of labeled features such as curb ramps, surface problems, obstructions, and crossings. We calculate Access Scores at three levels: street segment, POI, and POI-across-sector. 

\subsection{Accessibility Scoring}
Each labeled segment is represented by an accessibility feature vector $x_a$ segment. To account for severity, we assign each label a weight, forming a weight vector $w_a$ with values between 0.2 and 1.0, converted from severity ratings 1–3 (0.2 for severity 1, 0.6 for severity 2, and 1.0 for severity 3). For example, a surface problem with severity 3 (non-passable) receives a weight of 1.0. Higher weights indicate more severe conditions and thus lower accessibility. Positive features (such as curb ramps and marked crossings) increase the score, while negative ones (such as surface problems) reduce it. 

\subsubsection{Segment-level accessibility score (SegScore)} \label{subsubsec:SegScore}
The segment-level accessibility score is computed as the normalized dot product of the feature and significance vectors:
\[AS_{segment} = \frac{1}{1+e^{-w_s \cdot x_a}}\]


\subsubsection{POI-Level accessibility score (POISecScore)} \label{subsubsec:POISecScore}
Beyond single segments, we compute POI-level accessibility to see how access varies around key destinations. For each POI, we take all sidewalk segments within a 1 km radius in the same sector and compute a length-weighted average of their segment scores.
\[AS_{POI} = \frac{\sum_{i=1}^n AS_i \times L_i}{\sum_{i=1}^n L_i}\]
where $AS_i$ is the accessibility score of segment, $L_i$ is its length, and $n$ is the total number of segments around the POI

\subsubsection{POIs-across-sector accessibility score (POIScore)} \label{subsubsec:POIScore}
To estimate across-sector accessibility, we combine the POI-level scores. We weight each category by its number of POIs and take the weighted average to obtain the across-sector POI accessibility score:
\[AS_{across sector, poi} = \frac{\sum_{j=1}^n AS_{POI_j} \times N_{poi_j}}{\sum_{j=1}^m N_{poi_j}}\]
where $AS_{POI_j}$ is the accessibility score of POI $j$ and $N_{poi_j}$ is the count of POIs of that type across all the sectors.
  

\subsection{Analysis}

We evaluate two components of our system: (1) the effectiveness of VLM-generated mission guidance for supporting human annotation, and (2) the accessibility outcomes produced through our POI-centric analysis of Chandigarh. Together, these evaluations assess both the human–AI interaction layer and the urban accessibility insights enabled by the  tool.

\subsubsection{VLM-Assisted Mission Guidance Analysis}
To evaluate the quality of AI-generated guidance we rated 50 street segments with three annotators (R1–R3) on Likert (1–5) for relevance, accuracy, and usefulness. We report means and variation to show overall quality, Spearman to see if raters ranked segments the same way, and quadratic-weighted $\kappa$ to check reliable agreement on the Likert scale (crediting near misses). We computed basic stats (mean, SD, min, max) for each criterion. As shown in Table~\ref{tab:llm_descriptive} ratings were strong overall: relevance was highest with little disagreement, and accuracy and usefulness were also high but showed slightly more variation, implying a few guidance messages were less precise or helpful.
\begin{table}[H]
\small
\centering

\begin{tabular}{lccccc}
\toprule
\textbf{Criterion} & \textbf{Mean} & \textbf{SD} & \textbf{Min} & \textbf{Max} & \textbf{N} \\
\midrule
Relevance  & 4.97 & 0.26 & 2 & 5 & 150 \\
Accuracy   & 4.40 & 0.71 & 2 & 5 & 150 \\
Usefulness & 4.61 & 0.70 & 1 & 5 & 150 \\
\bottomrule
\end{tabular}
\caption{Descriptive statistics of human ratings for VLM-generated guidance.}
\label{tab:llm_descriptive}
\end{table}

\begin{figure*}[t]
  \centering
  \begin{subfigure}{0.4\textwidth}
    \includegraphics[width=\linewidth]{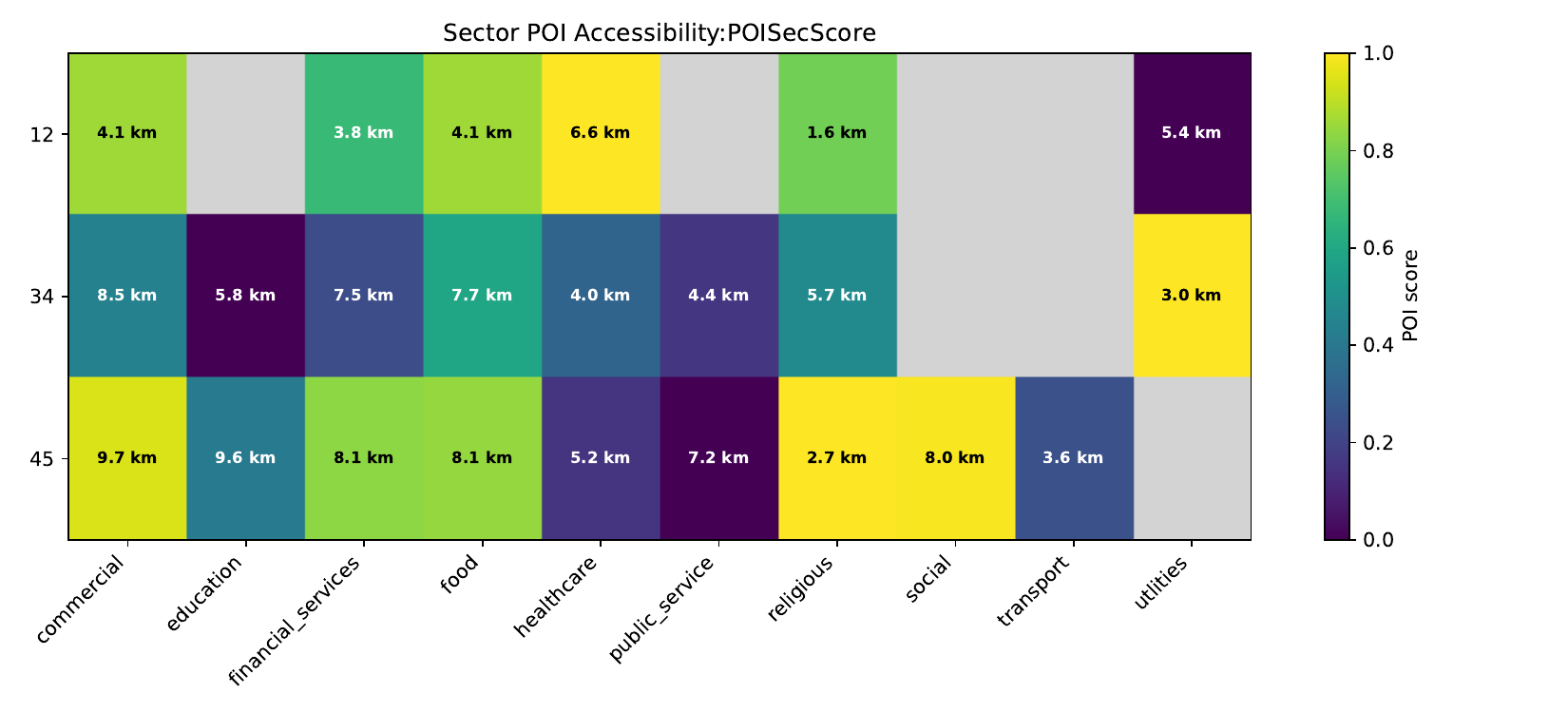}
    \caption{Heatmap of normalized POI accessibility scores.}
    \label{fig:poisecscore_heatmap}
  \end{subfigure}
  \begin{subfigure}{0.4\textwidth}
    \includegraphics[width=\linewidth]{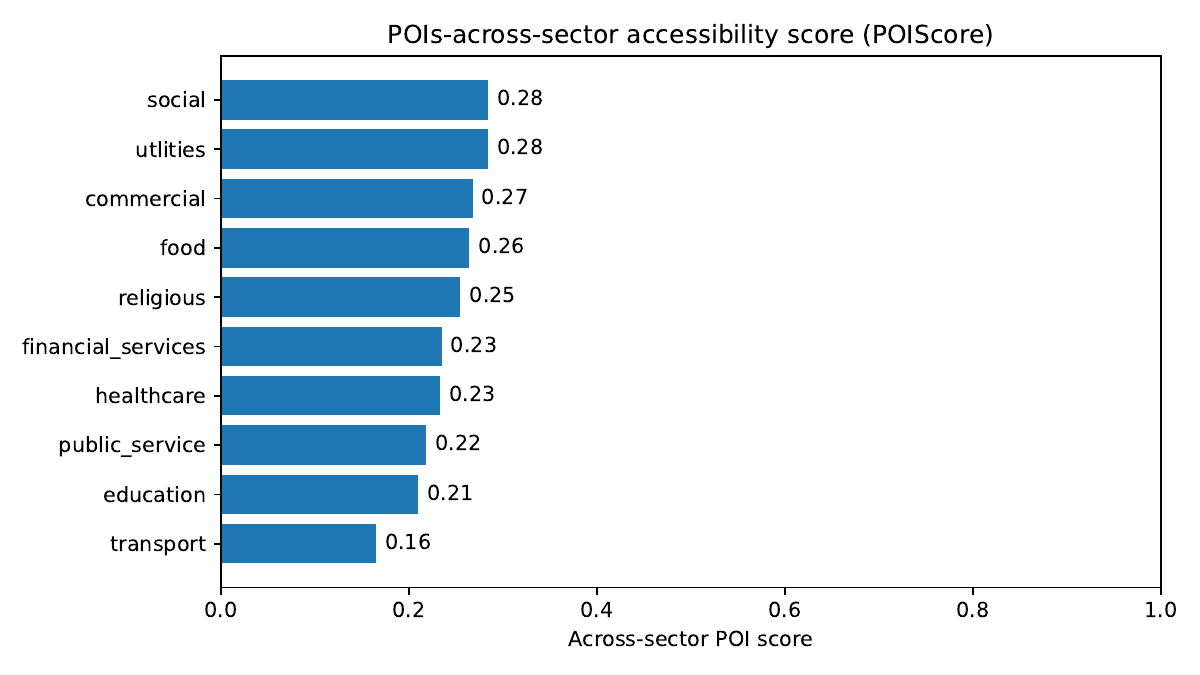}
    \caption{Across-sector bars.}
    \label{fig:poisecscore_bars}
  \end{subfigure}
  \caption{Sector POI Accessibility (POISecScore) and across-sector POI accessibility (POIScore). (a) Heatmap of sectors (rows) vs. POI categories (columns), (b) Category-wise summary of POI scores aggregated across sectors.}
  \label{fig:poisecscore_combo}
\end{figure*}

\begin{figure*}[t]
  \centering
  \includegraphics[width=0.3\textwidth]{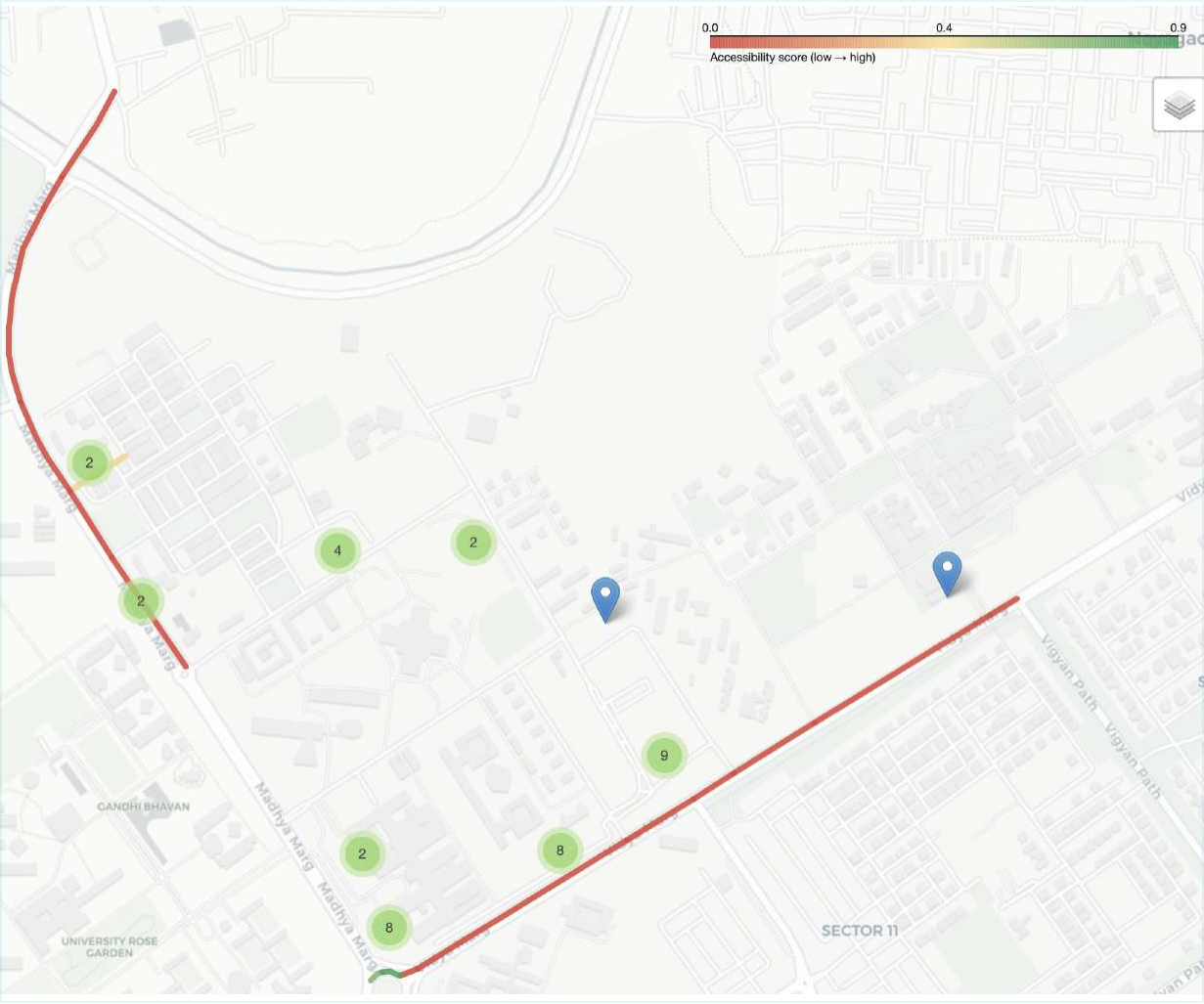}
  \includegraphics[width=0.3\textwidth]{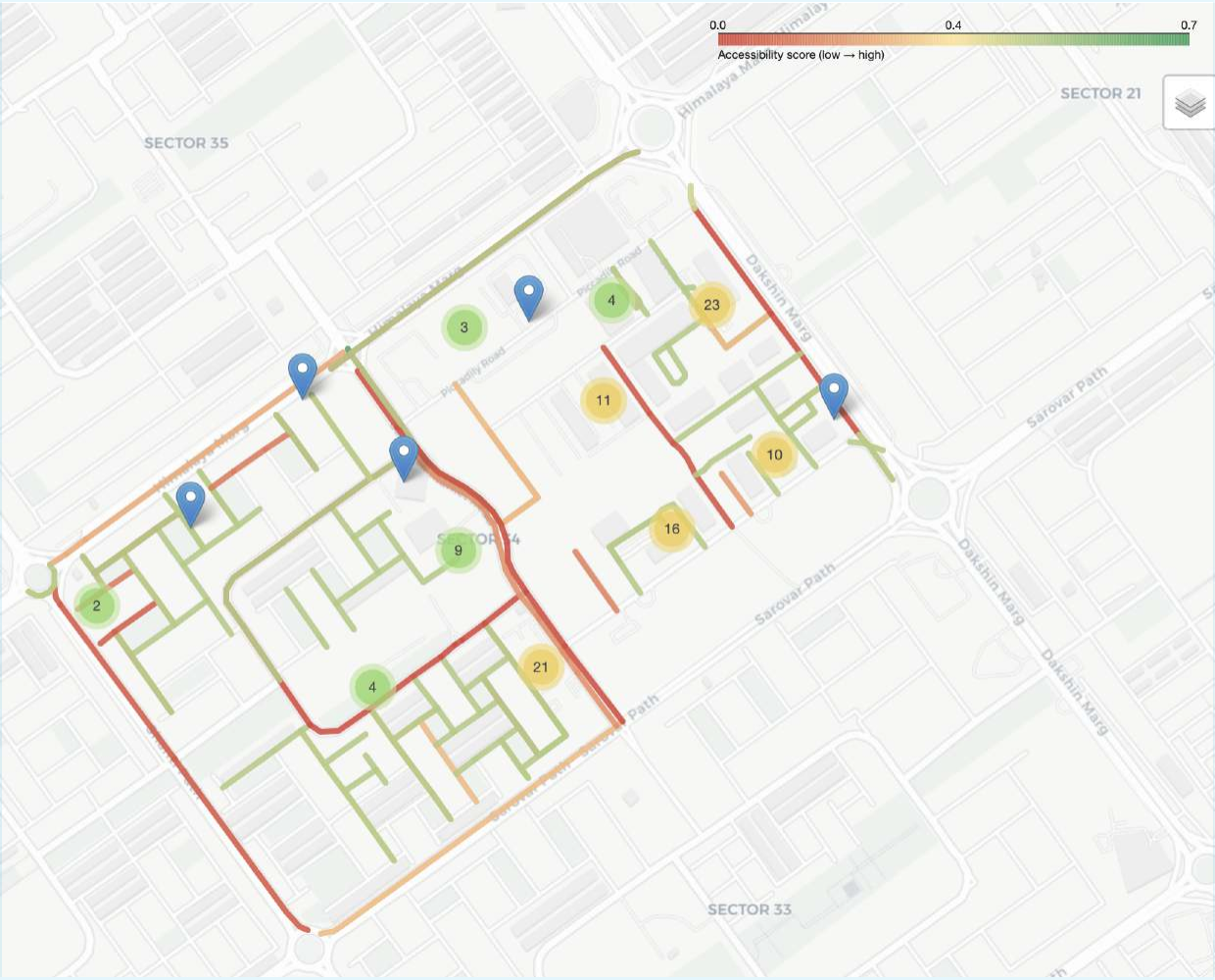}
  \includegraphics[width=0.3\textwidth]{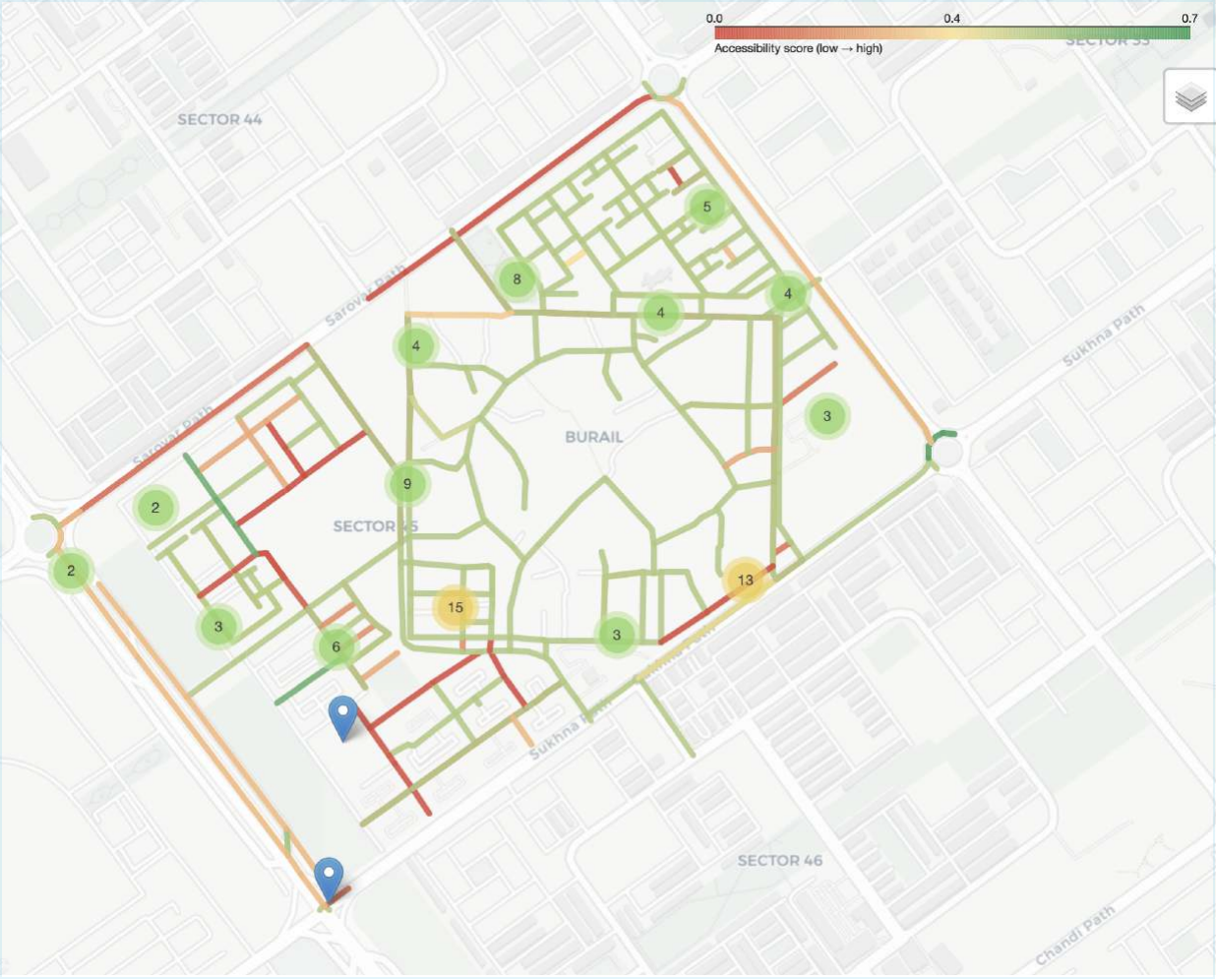}

  \caption{From left to right: SegScore heatmaps for sectors 12, 34, and 45 show the places of interest and the segscores for road segments. It highlights major road segments where accessibility improvements are required.}
  \label{fig:segmap}
\end{figure*}

We assessed rating consistency using pairwise Spearman correlations ($\rho$) among the three annotators. As shown in Table~\ref{tab:llm_spearman_kappa}, \textit{Relevance} was almost uniformly high (R1 and R3 were identical, \(\rho=1.0\)). For \textit{Accuracy}, R3 aligned moderately with the others ($\rho$ = 0.38-0.44), while R1 and R2 showed weaker alignment. For \textit{Usefulness}, only R2 and R3 showed a moderate match ($\rho$ = 0.45), suggesting similar judgments, whereas R1’s ratings differed slightly.


We assessed exact agreement with quadratic-weighted Cohen’s $\kappa$, which rewards near-matches on the 1–5 Likert scale. As shown in Table~\ref{tab:llm_spearman_kappa}, for \textit{Relevance} R1–R3 had perfect agreement (($\kappa$ = 1.0), while pairs with R2 were near chance (likely due to near-constant ratings). For \textit{Accuracy}, agreement was
moderate for R2–R3 ($\kappa$ = 0.49), fair for R1–R3 ($\kappa$ = 0.38), and weak for R1 R2. For \textit{Usefulness}, R2–R3 showed substantial agreement ($\kappa$ = 0.67), whereas R1’s agreement with others was lower. Overall, $\kappa$ indicates R2 and R3 were most consistent, especially on usefulness, while R1 varied more.
%
Overall, annotators rated the VLM guidance as highly relevant, mostly accurate, and generally useful. There was some variation on accuracy and usefulness, so the guidance can be made more precise. Raters mostly agreed with each other, suggesting the guidance was easy to understand and helpful for labeling.


\begin{table}[t]
\footnotesize       
\centering
\setlength{\tabcolsep}{3.5pt}  
\renewcommand{\arraystretch}{1.15}
\begin{tabular}{l c c c c}
\toprule
\textbf{Crit.} & \textbf{Pair} & \textbf{Spearman }$\rho$ & \textbf{$p$} & \textbf{Weighted $\kappa$} \\
\midrule
Relevance  & R1--R2 & \textit{n/a} & \textit{n/a} & 0.00 \\
Relevance  & R1--R3 & 1.000 & $<\!10^{-6}$ & 1.00 \\
Relevance  & R2--R3 & \textit{n/a} & \textit{n/a} & 0.00 \\
Accuracy   & R1--R2 & -0.209 & 0.146 & -0.113 \\
Accuracy   & R1--R3 &  0.384 & 0.0059 &  0.378 \\
Accuracy   & R2--R3 &  0.444 & 0.0012 &  0.487 \\
Usefulness & R1--R2 & -0.149 & 0.300 &  0.031 \\
Usefulness & R1--R3 &  0.019 & 0.895 &  0.183 \\
Usefulness & R2--R3 &  0.445 & 0.0012 &  0.665 \\
\bottomrule
\end{tabular}
\caption{Pairwise Spearman correlation and quadratic-weighted Cohen’s $\kappa$ for human ratings of VLM-generated guidance. ``n/a'' indicates non-computable correlations due to constant ratings.}
\label{tab:llm_spearman_kappa}
\end{table}

\subsubsection{POI analysis}
At each panorama, we marked `no sidewalk' only when that spot’s sidewalk was disrupted, broken, or missing---this is a local label and does not mean the whole path lacks a sidewalk. Raw scores had a long negative tail, so we clipped negative outliers at 95th percentile, standardized (mean 0, var 1), and applied a sigmoid. We then computed segment scores (SegScore); maps are in Figure~\ref{fig:segmap}, where 0 is least accessible and 1 is most accessible. Blue pins show POIs from 10 categories; numbered circles are POI clusters. 

Figure~\ref{fig:poisecscore_combo}a shows Sector POI Accessibility (normalized computed using the POISecScore equation. It plots sectors (rows) against POI categories (columns), with color is POI score (0–1) and in-cell text indicates total road length (km) for that category in the sector. Gray cells indicate categories not available in a given sector (no data). Brighter colors denote more accessible conditions. Figure~\ref{fig:poisecscore_combo}b summarizes the category-wise POI scores across sectors.

Sector 45 shows the highest scores for residential, religious, social, and commercial POIs; other categories need attention. In Sector 12 (institutional), accessibility is best around healthcare, but other categories especially utilities are weak and need work. Sector 34 (commercial) shows moderate access for commercial POIs, with more improvement needed in other categories. Across sectors, education and transport score the lowest and should be prioritized.



\section{Discussion and Future work}
We adapted Project Sidewalk for Chandigarh, India by contextualized labels, example images, and a VLM-based mission guidance. A three-user test rated the guidance 4.66/5, indicating practical usefulness. Using the adapted tool, we audited about 40 km of streets across three sectors and around 230 POIs, and found 1,644 locations where fixes could improve access. In our results, commercial areas show better access overall, while education and public-service areas are weaker; in the institutional sector, healthcare is relatively accessible but everyday places like transport stops and food outlets are not. This marks the start of a journey for accessibility mapping in multiple Indian cities.

\section{Acknowledgments}
This work is partially supported by the Indorama Ventures Center for Clean Energy at Plaksha University and NSF \#2125087 at the University of Washington. Ms Varchita Lalwani is a recipient of the Harsh and Bina Shah School of Computer Science and Artificial Intelligence.

\bibliography{aaai2026}


\end{document}